\documentclass[utf8]{FrontiersinHarvard}
\usepackage{url,microtype,subcaption}
\usepackage[hidelinks]{hyperref}
\usepackage[onehalfspacing]{setspace}
\usepackage{booktabs}
\usepackage{graphicx}

\def\keyFont{\fontsize{8}{11}\helveticabold }
\def\firstAuthorLast{Liu {et~al.}}
\def\Authors{Pufan Liu\,$^{1,2}$, Hui Li\,$^{1}$, Ziqi Li\,$^{1,2}$, Xiaoyue Cao\,$^{3,4,1}$, Rui Li\,$^{1*}$, Hao Su\,$^{5}$, Ran Li\,$^{6}$, Nicola R. Napolitano\,$^{7}$, L\'{e}on V. E. Koopmans\,$^{8}$,Valerio Busillo\,$^{9}$, Crescenzo Tortora\,$^{9}$, Liang Gao\,$^{1,6}$}

\def\Address{
$^{1}$Institute for Astrophysics, School of Physics, Zhengzhou University, Zhengzhou, 450001, China \\
$^{2}$International College, Zhengzhou University, Zhengzhou, 450001, China \\
$^{3}$School of Astronomy and Space Science, University of Chinese Academy of Sciences, Beijing 100049, China \\
$^{4}$National Astronomical Observatories, Chinese Academy of Sciences, 20A Datun Road, Chaoyang District, Beijing 100012, China\\
$^{5}$Department of Physics “E. Pancini”, University of Naples Federico II, Via Cintia, 21, 80126 Naples, Italy\\
$^{6}$ School of Physics and Astronomy, Beijing Normal University,  Beijing 100875, China\\
$^{7}$Department of Physics ``E. Pancini'', University Federico II, Via Cinthia 6, 80126-I, Naples, Italy\\
$^{8}$Kapteyn Astronomical Institute, University of Groningen, P.O.Box 800, 9700AV Groningen, the Netherlands\\
$^{9}$INAF -- Osservatorio Astronomico di Capodimonte, Salita Moiariello 16, 80131 - Napoli, Italy
}

\def\corrAuthor{Corresponding Author}

\def\corrEmail{liruiww@gmail.com}

\begin{document}
\onecolumn
\firstpage{1}
\title[Scalable Lens Finding with LenNet]{LenNet: Direct Detection and Localization of Strong Gravitational Lenses in Wide-Field Sky Survey Images}

\author[\firstAuthorLast ]{\Authors} 
\address{} 
\correspondance{} 

\extraAuth{}

\maketitle
\begin{abstract}
Strong gravitational lenses are invaluable tools for addressing fundamental questions in astrophysics, from the nature of dark matter to the expansion of the universe. While current sky surveys have successfully identified thousands of lens candidates, the search methods employed face a critical challenge. The conventional approach relies on a ``crop-and-classify'' strategy, where small images are first cut out around billions of potential host galaxies before being individually classified. This process creates a significant computational and storage bottleneck that is unsustainable for future large-scale surveys. To overcome this limitation, we propose LenNet, an object detection model that identifies lenses directly within large, original survey images. Our method completely bypasses the inefficient cropping step by framing the problem as a direct detection and localization task. We initially train LenNet on simulated data to learn the complex features of gravitational lenses and then use transfer learning to fine-tune the model on a limited set of real, labeled examples from the Kilo-Degree Survey (KiDS). Our experiments show that LenNet performs remarkably well on real survey data, validating its potential as a highly efficient and scalable solution for lens discovery in massive astronomical surveys.

\tiny
 \keyFont{ \section{Keywords:} Machine Learning, Gravitational Lensing, Object Detection} 
\end{abstract}

\section{Introduction}
According to Einstein's Theory of General Relativity, light from a distant source galaxy travels to the observer along geodesics in spacetime. If a foreground galaxy is closely aligned with such a source, its gravitational potential distorts the intervening spacetime, deflecting the light to form multiple images or extended arcs \citep{Schneider1992_book}. This phenomenon, analogous to optical lensing, is known as galaxy-galaxy strong gravitational lensing (GGSL) and constitutes a powerful astrophysical tool \citep{Shajib2024_review, Treu2010_review}. GGSL provides robust estimates of the lens galaxy’s mass distribution \citep{Koopmans2006_mass, Auger2010_mass, Sonnenfeld_2015, Shajib2021, sheu2025project}, including its dark matter subhalos \citep{Vegetti2012, Nightingale2024, Ballard2024_0946, Cao2025, He2025}; when combined with stellar population synthesis models, these measurements yield constraints on the galaxy’s initial mass function \citep[e.g.][]{Sonnenfeld_2015, Li2025}. Furthermore, GGSL acts as a natural `cosmic telescope', magnifying background sources to enable the study of their structure \citep{Dye15, Shu16, Cheng20, Rizzo20, Li2024}, which would otherwise be beyond the capabilities of current instruments. Finally, because light propagates through the cosmological background, GGSL serves as an independent cosmographic probe when the background source exhibits temporal variability \citep{Treu2022, Birrer2024} or when multiple sources lie at different redshifts \citep{Gavazzi08, Collett14, Litian_2025}.

GGSL is a rare astrophysical event that necessitates precise alignment, typically within a few arcseconds, between foreground and background galaxies. Consequently, identifying individual lenses within contemporary sky surveys often requires extensive visual inspection of numerous galaxy images, rendering the development of automated detection methods imperative, particularly as forthcoming surveys will catalogue more than billions of galaxies \citep[e.g.][]{LSST_galaxy_counts}. Traditional automatic lens-finding approaches include spectrum-based techniques \citep{Bolton_2006, Shu_2016, Cao_2020}, which identify anomalous emission lines in foreground galaxies indicative of background sources, and image-based methods \citep{Cabanac_2006, More_2012, nightingale2025cosmos}, which directly inspect galaxy images for characteristic lensing features such as bluish arcs. While these conventional methods have successfully confirmed several hundred GGSLs, recent advances in deep learning, particularly convolutional neural networks (CNNs), have significantly enhanced lens discovery capabilities, yielding thousands of new candidates \citep[e.g.][]{Petrillo_2017_lens_find, Jacobs2019_lens_find, Li2021_lens_find, Nagam2025_lens_find}. Nevertheless, existing lens samples are still insufficient for many strong lens sciences in terms of size and completeness \citep{Shajib2024_review}. Ongoing and Upcoming telescope facilities, including Euclid \citep{Euclid_lens_find}, the China Space Station Telescope \citep[CSST,][]{CSST_zhanhu}, the Nancy Grace Roman Space Telescope \citep[Roman,][]{Roman}, and the Vera C. Rubin Observatory's Legacy Survey of Space and Time \citep[LSST,][]{LSST}, are expected to detect hundreds of thousands of GGSLs \citep{Collett15, Cao2024, Wedig2025}. This anticipated surge in available data promises to revolutionise the field, but concurrently underscores the critical need to develop more efficient and robust automated lens-finding algorithms to fully exploit this unprecedented opportunity.

The conventional pipeline for identifying gravitational lenses using neural networks typically comprises three main stages. First, potential lens candidates, such as massive elliptical galaxies, are pre-selected from a parent source catalogue generated by software such as Source Extractor \citep{SExtracto}. Second, small cutout images centred on these candidates are produced. Third, a neural network, commonly a CNN, classifies each cutout to determine if it contains a strong lens. This methodology has been extensively employed in various large-scale surveys \citep[e.g.][]{Petrillo2019, Li2021_lens_find, Euclid_lens_find}, achieving notable success. However, this approach has several limitations. Firstly, in crowded fields, the source-detection algorithm may fail, leading to cutouts that blend multiple neighbouring objects. Such blending can hinder the network's ability to accurately recognise genuine lensing features. Secondly, choosing an appropriate cutout size involves a critical trade-off that directly influences detection accuracy. If the cutout is too small, lensed arcs of systems with large Einstein radii may lie outside the field of view, resulting in missed detections. Conversely, overly large cutouts may include nearby contaminating objects, especially in crowded regions, potentially causing the network to misidentify these contaminants as lensing features, thus generating false positives. Thirdly, the cutout-based approach inherently neglects information about the broader environmental context of the lens system. This context can provide subtle yet valuable clues for lens identification; for example, GGSLs are more likely to occur within dense galaxy cluster environments \citep{Oguri_2005}. Finally, the pre-selection stage can systematically exclude rare yet scientifically interesting systems, such as those with very low mass \citep{Shu_2017} or disk lensing galaxies \citep{Treu_2011}.

To address the limitations inherent in conventional neural-network-based gravitational lens finders,
we introduce \texttt{LenNet}, a neural network specifically designed to identify GGSLs directly from full-survey images. This approach eliminates the necessity of pre-selecting target sources and extracting individual image cutouts\footnote{\cite{Lixu_2024} also employ a transformer-based neural network to directly search for GGSLs in full-survey images. However, their method still assumes a fixed `window size', potentially missing lenses whose Einstein radius exceeds this predefined size.}. \texttt{LenNet} incorporates advanced feature-extraction techniques explicitly tailored for gravitational lens identification. Initially trained on simulated lensing datasets, \texttt{LenNet} subsequently employs transfer learning to improve its generalisation capabilities for application to real observational data. On mock lensing datasets, \texttt{LenNet} achieves a precision of 99.89 per cent with a recall rate of 98.80 per cent, while on real KIDS lensing datasets, it attains an accuracy of 97.50 per cent with a recall rate of 95.12 per cent. Additionally, we compare the performance of \texttt{LenNet} with several classical object detection neural networks (e.g., YOLO\citep{khanam2024yolov5,redmon2018yolov3}) and find that \texttt{LenNet} demonstrates superior performance.

This paper is structured as follows. Section~\ref{sec:2} describes the methodology used to generate the simulated dataset. Section~\ref{sec:3} details the architecture, key components, and training procedure of \texttt{LenNet}. Section~\ref{sec:4} presents the results, including performance evaluations of \texttt{LenNet} on both simulated and real observational data. Finally, Section~\ref{sec:conclusion} summarises our conclusions. 

\section{Dataset generation}\label{sec:2}

\subsection{KIDS DR4}
The Kilo-Degree Survey (KiDS) is a Stage-III optical imaging survey conducted using the OmegaCAM wide-field camera on the VLT Survey Telescope (VST) at the ESO Paranal Observatory \citep{Kuijken_2015}. Initiated in 2011, KiDS was designed to map a large area of the extragalactic sky with exceptional image quality. Although its primary scientific goal is weak gravitational lensing for cosmological studies, the survey's deep, multi-band imaging data provide a rich resource for a broad range of astrophysical investigations, including strong lensing, galaxy evolution, and the characterisation of galaxy clusters. The KiDS footprint covers 1350\,$\text{deg}^2$, divided into two large patches on the North and South Galactic Caps. The fourth data release (DR4) encompasses 1006\,$\text{deg}^2$ of observations in four broad-band optical filters---$u$, $g$, $r$, and $i$ \citep{Kuijken_2019}. To enable precise galaxy shape measurements, observations in the $r$-band were prioritised for optimal seeing conditions, requiring a point-spread function (PSF) with a full-width at half-maximum (FWHM) of less than $0.8^{\prime\prime}$. The median seeing achieved in the $r$-band is $\approx 0.7^{\prime\prime}$. Furthermore, the instrument optics ensure a uniform PSF across the $1\,\text{deg}^2$ field of view, establishing KiDS as one of the sharpest large-area, ground-based surveys. The survey is notably deep, reaching $5\sigma$ limiting magnitudes in a $2^{\prime\prime}$ aperture of $u \approx 24.2$, $g \approx 25.1$, $r \approx 25.0$, and $i \approx 23.7$. This depth, when combined with near-infrared data from the VISTA Kilo-Degree Infrared Galaxy (VIKING) survey \citep{Edge_2013}, allows the galaxy population to be probed to a median redshift of $z \approx 0.8$ and beyond.

\subsection{Lens Simulation}\label{sec:lens_simulation} 
Identifying GGSLs using neural networks requires substantial amounts of training data. However, confirmed lenses remain scarce within the KiDS survey\citep{Petrillo2019, 2020ApJ...899...30L, 
Li2021_lens_find, 
grespan2024teglie, 2024MNRAS.533.1426N}. Therefore, we have developed a specialized GGSL simulation pipeline to generate realistic training samples. This method, described extensively in \citet{Li2021_lens_find}, involves injecting simulated lensing images into real KiDS galaxy observations, as detailed below.

It should be noted that lens-free images are not essential for the model training and validation in this study. On one hand, if lens-free images were to be used for model optimization, they would first require manual visual inspection to screen out valid samples that are free of interference and errors. This process consumes significant labor and time costs, thereby reducing research efficiency. On the other hand, the core objective of this study is to improve the model's accuracy in recognizing "lenses". Importantly, lens-free images do not actually participate in the calculation of loss; their auxiliary role in helping the model learn the "key features of lenses" is extremely minimal. Even without incorporating lens-free images, the model can still maintain excellent detection capabilities.

Luminous Red Galaxies (LRGs) are selected from the KiDS DR4 catalogue to act as potential lens galaxies. Following the methodology of \citet[Section 3.1]{Li2021_lens_find}, the selection is based on both colour and magnitude criteria, with a magnitude limit of $r < 20.0$  and colors satisfying a slightly adapted LRG color–magnitude condition.
This imposes to selection of massive galaxies, which are the most efficient deflectors\footnote{Due to this choice, the \texttt{LenNet} neural network trained in this study is optimised to find GGSLs with LRG deflectors. Nevertheless, \texttt{LenNet}'s capabilities could be extended to identify GGSLs with disc or low-mass galaxy deflectors by introducing greater diversity and realism into the training set generation.}. 
The mass distribution of each LRG is modelled using a singular isothermal ellipsoid (SIE) profile, supplemented by an external shear component to account for environmental effects. The Einstein radius $\theta_E$, ellipticity, and shear strength are randomly drawn from distributions specified in \citet{Li2021_lens_find}. Background sources are constructed using morphological and colour information from the CosmoDC2 simulation catalogue with selected redshifts between 0.8 and 6 \citep{CosmosDC2} . Each randomly selected source is then strongly lensed by a prospective foreground LRG within a large $501 \times 501$ pixels ($0.2^{\prime\prime}/ \mathrm{pixels}$, corresponding to $100.2^{\prime\prime}\times 100.2^{\prime\prime}$) survey image to generate an idealised lensed image. The position of this prospective deflector is randomly assigned within the survey image. Specifically, we select sources that satisfy the criterion $z_\mathrm{source}>z_\mathrm{lens}+0.3$.  To replicate the spatial resolution of real observations, this idealised lensed image is subsequently convolved with the local point spread function (PSF) extracted from the corresponding region of the KiDS imaging data. The PSF-convolved lensed image is then injected into the KiDS survey image, centred on the prospective foreground LRG, thereby creating a realistic training sample. For the object detection task, the precise location of the injected lens must be recorded. We determine the coordinates of the lensed source by defining a bounding box that tightly encloses the full extent of the simulated lensed arcs or ring. These bounding box coordinates, along with the class label indicating a lens, constitute the ground truth for training the object detection model.

\begin{figure}[h!]
    \begin{center}
        \includegraphics[width=0.9\textwidth]{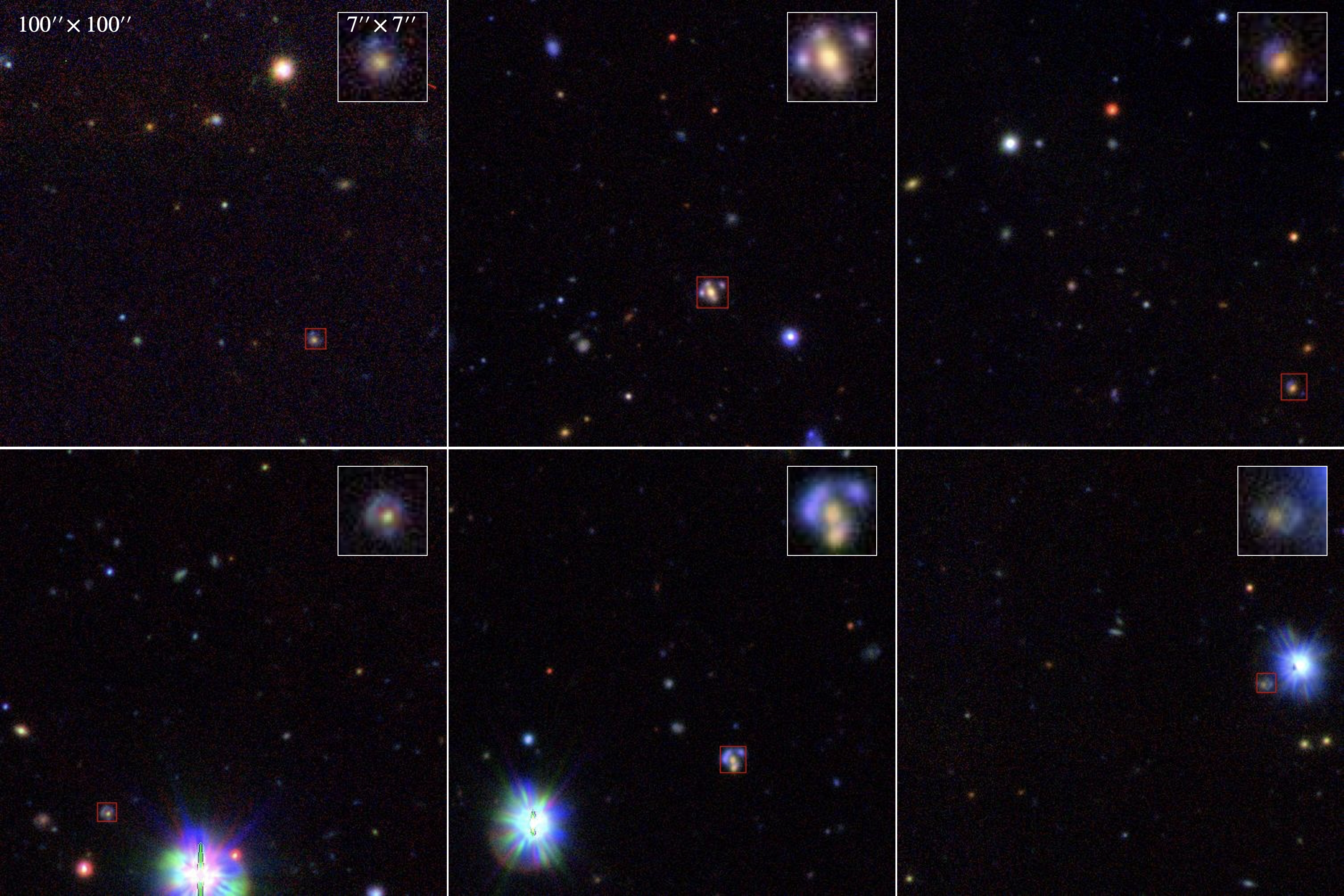}
    \end{center}
    \caption{Representative examples of simulated strong lenses in KiDS DR4 imaging. Each system is marked by a red box, and the inset shows a zoomed-in view of the lens.} 
    \label{fig: mock lenses}
\end{figure}

This simulation pipeline yielded 9,147 mock lenses with known labels for network training. 
Notably, one simulated field contains only a single lens. The full sample of lenses was simulated with Einstein radii in the range of 1--5~arcsec, lens redshifts ($z_{\text{lens}}$) from 0.08 to 0.80, and source redshifts ($z_{\text{source}}$) from 0.80 to 4.84.
Figure \ref{fig: mock lenses} presents six representative examples, illustrating the complexity and realism achieved by our simulations. These simulations successfully reproduce the diverse morphologies of strong lenses, whilst preserving the noise characteristics, PSF effects, and galaxy populations typical of KiDS observations.

\section{Method}\label{sec:3}
The performance of LenNet is determined by three core components: its model architecture, loss function, and training procedure. The architecture and loss function define the model's theoretical performance ceiling, while the training strategy determines how effectively this potential is achieved. The following subsections detail each of these three components.

\subsection{Network architecture}

\begin{figure}[h!]
    \begin{center}
        \includegraphics[width=15cm]{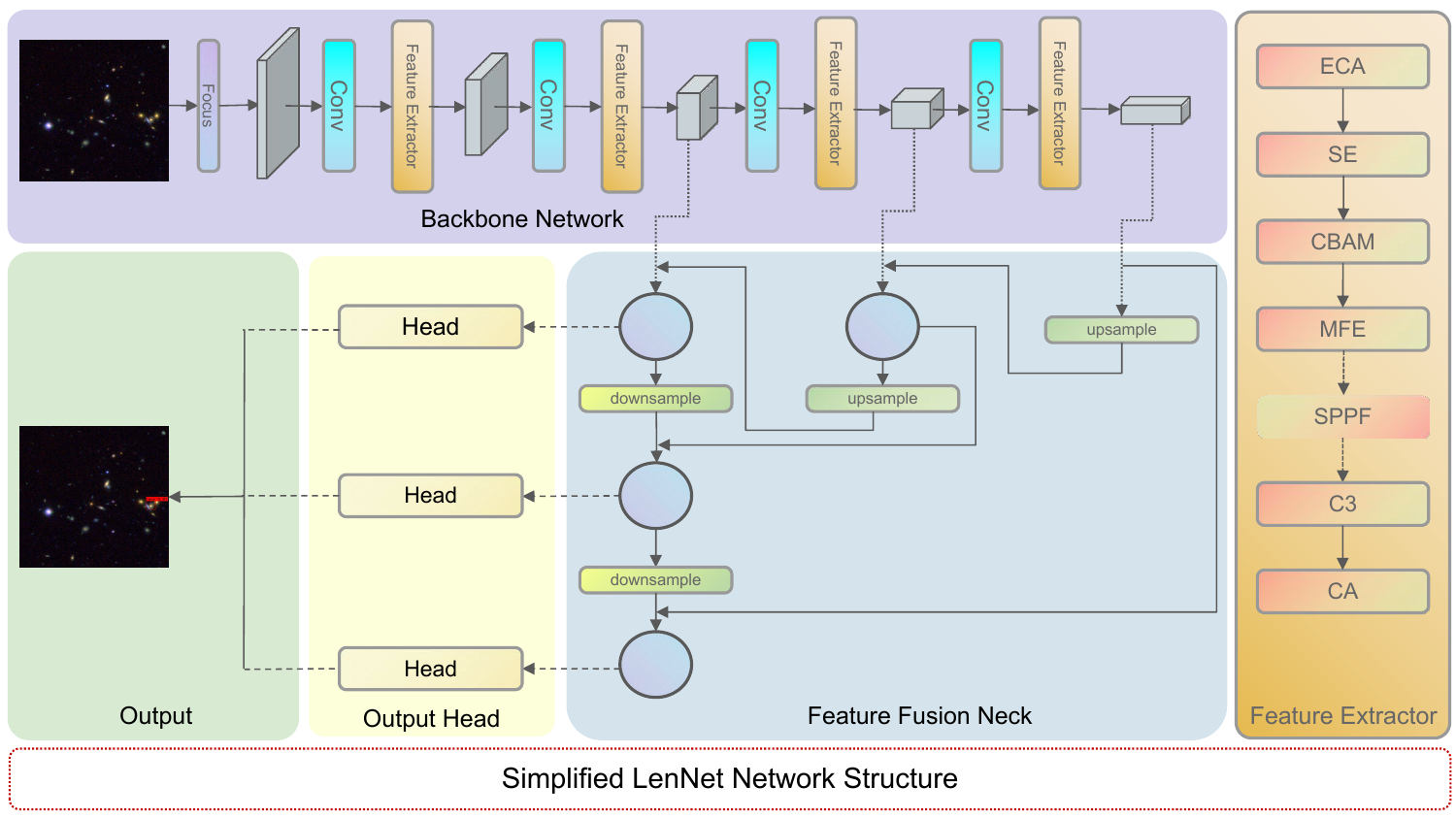}
    \end{center}
    \caption{Simplified LenNet Network Structure. This figure is intended solely to provide readers with a general understanding of the overall architecture of LenNet, and does not encompass all components in detail. For example, in the upsampling and downsampling process of the feature fusion neck(i.e., FPN Network), the intermediate feature extraction layers interleaved between these operations have been omitted for clarity. In the diagram, dashed lines represent the connections between different components, with rectangular blocks in the Backbone Network and circular nodes in the Feature Fusion Neck both denoting intermediate feature maps.} 
    \label{fig:Model Structure}
\end{figure}

Figure \ref{fig:Model Structure} provides a simplified schematic of this architecture, omitting intermediate layers for clarity. The model's architecture is composed of two primary stages: a feature extraction backbone and a feature fusion neck. These modules are followed by a detection head, which processes the fused feature maps at three distinct scales. This multi-scale approach enables the model to effectively detect gravitational lenses of various apparent sizes. The operational workflow diverges for training and inference. During the training phase, the model's predictions are compared against ground-truth labels to compute a loss value, which guides parameter updates via backpropagation. Conversely, during inference, the raw predictions are aggregated and refined using Non-Maximum Suppression (NMS)\citep{neubeck2006efficient} to produce the final set of detections.

The feature extraction module serves as the backbone of LenNet, designed to transform raw input images into a hierarchy of representative and discriminative features. This is achieved by progressively downsampling the spatial dimensions while extracting information critical for lens identification. The module's architecture is a carefully orchestrated sequence of attention mechanisms, a bespoke Multi-Feature Extraction block (MFEBlock\citep{xie2023shisrcnet}), and the efficient C3 module from CSPNet \citep{wang2023yolov7,khanam2024yolov5}. It starts with a lightweight ECA block \citep{wang2020eca} for initial channel-wise weighting. Following this, an SE block \citep{hu2018squeeze} further refines these channel weights, focusing computational resources on the most informative channels. Building on this, a CBAM block \citep{woo2018cbam} introduces a spatial attention component, creating a comprehensive feature map refined in both channel and spatial domains. After this attention-based refinement, the core feature extraction is performed. The MFEBlock employs dilated convolutions to systematically enlarge the receptive field, enabling the capture of broad contextual information essential for identifying lens systems. The subsequent C3 module then effectively aggregates the multi-scale features generated by the MFEBlock, utilizing residual connections and feature concatenation to produce a semantically rich representation. Finally, the sequence concludes with a Coordinate Attention (CA) block \citep{hou2021coordinate} to capture direction-aware and position-sensitive information, which is crucial for the fine-grained localization of the target lens. In the final feature extraction stage only, an SPPF block \citep{he2015spatial} is appended to fuse features at multiple scales efficiently. The output of this entire module is then passed to the next stage of the backbone or serves as one of the three final feature maps for the detection head. The unique presence of the SPPF block in the last stage is highlighted in Figure \ref{fig:Model Structure}.

In the feature fusion neck section, to effectively integrate features across different scales, we employ a feature fusion neck based on the Path Aggregation Network (PANet) architecture, a well-established enhancement to the classical Feature Pyramid Network (FPN) \citep{lin2017feature}. This neck structure creates bidirectional information flow, combining a top-down path to propagate rich semantic features with a bottom-up path to preserve precise localization information. Let P3, P4, and P5 denote the multi-scale feature maps produced by the backbone. Prior to fusion, each map is passed through an SE block to recalibrate channel-wise feature responses. The process then unfolds in two stages:
\begin{itemize}
    \item Top-Down Pathway: This path propagates high-level semantic information from deeper layers (P5) to shallower ones. Specifically, the P5 feature map is upsampled and fused (e.g., via concatenation) with P4. This combined map is then processed and upsampled again to be fused with P3, creating a set of semantically enhanced feature maps.
    \item Bottom-Up Pathway: Following the top-down pass, a complementary bottom-up pathway is constructed. This path transmits low-level, high-resolution positional information from the shallower layers (e.g., the newly enhanced P3) upwards to the deeper layers. This is achieved through a series of downsampling and fusion operations, ensuring that the final output pyramids contain both strong semantic and precise spatial information.
\end{itemize}
The resulting feature maps from this bidirectional fusion process—P3, P4, and P5—are information-rich and serve as the final inputs to the detection head for generating predictions.

\subsection{Loss function}

The overall loss function, $L_{total}$, is a composite objective designed to train the model on three distinct tasks: objectness confidence prediction, class prediction, and bounding box localization. It is a weighted sum of a confidence loss ($L_{conf}$), a classification loss ($L_{cls}$), and a localization loss (
$L_{loc}$). Both the confidence and classification losses are implemented using the standard Binary Cross-Entropy (BCE) loss. The localization loss is calculated using the Generalized Intersection over Union (GIoU) \citep{rezatofighi2019generalized}, which offers more stability than the standard IoU for non-overlapping boxes.
The individual loss components are defined as follows:
\begin{equation}
L_{\text{conf}} = -\frac{1}{N} \sum_{i=1}^{N} \left[ y_i^{\text{conf}} \cdot \ln(p_i^{\text{conf}}) + (1 - y_i^{\text{conf}}) \cdot \ln(1 - p_i^{\text{conf}}) \right]
\label{eq:conf_loss_corrected}
\end{equation}

\begin{equation}
L_{\text{cls}} = -\frac{1}{N} \sum_{i=1}^{N} \left[ y_i^{\text{cls}} \cdot \ln(p_i^{\text{cls}}) + (1 - y_i^{\text{cls}}) \cdot \ln(1 - p_i^{\text{cls}}) \right]
\label{eq:cls_loss_corrected}
\end{equation}

\begin{equation}
L_{\text{loc}} = L_{\text{GIoU}} = 1 - \left( \text{IoU} - \frac{|C| - |A \cup B|}{|C|} \right)
\label{eq:GIoU_corrected}
\end{equation}
where $N$ is the number of predictions. $y_i^{conf}$ and  $p_i^{conf}$ are the ground-truth and predicted confidence scores for the i-th prediction, respectively.
$y_i^{cls}$ and $p_i^{cls}$ are the ground-truth and predicted class probabilities. Intersection over Union (IoU) is the Intersection over Union between the ground-truth box $A$ and the predicted box $B$. $C$ is the area of the smallest convex bounding box enclosing both $A$ and $B$.
The final objective function combines these components, weighted according to their importance for the single-class lens detection task. Since accurate localization is the primary goal, we prioritize the confidence and localization losses. Following a similar strategy to LSGBnet \citep{su2024lsbgnet}, the classification loss is assigned a minimal weight, mainly serving to stabilize training in the multi-task framework. The final loss is:
\begin{equation}
\mathcal{L}_{\text{total}} = \omega_{\text{conf}} L_{\text{conf}} + \omega_{\text{cls}} L_{\text{cls}} + \omega_{\text{loc}} L_{\text{loc}}
\end{equation}
where the weights are set to $\omega_{conf}=10$, $\omega_{cls}=0.005$, and $\omega_{loc}=1$.

\subsection{Training}
In object-detection frameworks like ours, the training data already includes negative examples via the background regions of images that contain lenses. With enough training images, these backgrounds provide ample negatives, allowing the model to learn discriminative features without additional pure negative images. This follows standard practice in modern object detection networks (e.g., YOLO, Faster R-CNN), which typically do not train on entire negative images.

In our simulation-based pretraining, we supply many simulated lens images, whose background regions already yield numerous negative samples. Adding pure negative images would likely not improve performance and would introduce practical challenges: verifying that large numbers of candidate negatives truly contain no lenses is labor-intensive and error-prone, especially at the scale of the simulated dataset. However, during transfer learning on real data (see Section \ref{sec:4 trans}), where the training set is much smaller (\(\sim 200\)–\(360\) images), we included a small set of carefully verified negative samples. This improves the model’s ability to distinguish lenses from specific background features in real observations, and manual verification is feasible at this scale.

We generated a comprehensive dataset of simulated gravitational lens images for this study. The complete dataset was partitioned into training, validation, and test sets following a standard 8:1:1 ratio. All splits were sampled from the same underlying data distribution to ensure an unbiased evaluation of the model's final performance.

The model was trained using the Adam optimizer \citep{kingma2014adam} with a batch size of 32. The initial learning rate was set to \ensuremath{2 \times 10^{-5}} and was managed by a ReduceLROnPlateau scheduler, which halved the learning rate whenever the validation loss stagnated for five consecutive epochs.We conducted our training process using a single NVIDIA A100 GPU, leveraging the PyTorch deep learning framework (version 2.4.1) with Python 3.12.7 and CUDA version 12.0. The training spanned 150 epochs and required a total of 9 GPU hours. Throughout the training, the GPU achieved an average utilization rate of 95\%, while the memory and CPU utilization rates averaged 36\% and 67\%, respectively. Additionally, during the testing phase, the test dataset comprised 915 images containing lenses and 1200 images without lenses, and LenNet took only 28 seconds to make predictions on all of them.It is particularly worth noting that these 1200 lens-free images are real-world samples selected completely at random.

\section{Results and discussion}\label{sec:4}

To comprehensively assess the performance of LenNet, we employ a combination of quantitative metrics and qualitative analysis. For the quantitative evaluation, we adopt three standard metrics from the field of object detection: Precision, Recall, and the F1-score. The Recall measures the model's ability to identify all actual gravitational lenses within the dataset, thus quantifying its completeness. Precision measures the reliability of the predictions, indicating what fraction of the identified candidates are actual lenses. A fundamental trade-off exists between these two metrics; for instance, a model can achieve high recall by lowering its detection threshold, but this often leads to an increase in false positives and a decrease in precision. The F1-score, which is the harmonic mean of precision and recall, provides a single, balanced metric that summarizes the model's overall accuracy. These metrics are defined as:
\begin{align*}
\text{Recall} &= \frac{TP}{TP + FN}, \\
\text{Precision} &= \frac{TP}{TP + FP}, \\
\text{F1-score} &= \frac{2 \times \text{Precision} \times \text{Recall}}{\text{Precision} + \text{Recall}}.
\end{align*}
Where TP is the True Positive, means that a predicted bounding box has an IoU with a ground-truth box that exceeds a predefined threshold (e.g., $\rm{IoU} > 0.5$). FP (False Positive) corresponds to a predicted box with no matching ground-truth object, while a FN (False Negative is a ground-truth object that the model failed to detect. 

Beyond these numerical scores, it is crucial to understand how the model arrives at its predictions. To this end, we perform a qualitative analysis by visualizing the model's internal feature maps. This interpretability technique allows us to peer inside the model's ``black box'' and gain direct, intuitive insight into its decision-making process. Specifically, this analysis enables us to verify that LenNet learns to activate on physically meaningful structures—such as Einstein rings, arcs, or multiple images—rather than relying on spurious artifacts or background noise. This confirmation is vital for building trust in the model's scientific utility and ensuring that its high performance is rooted in a genuine understanding of the underlying astrophysics.

\subsection{The performance on the simulated dataset}
\label{sec:mock_perform}
This section details the quantitative performance of LenNet on our simulated dataset. We first characterize the model's standalone effectiveness using standard evaluation metrics and then contextualize these results through a rigorous comparative analysis against established, state-of-the-art object detection models.

\begin{figure}[h!]
    \centering 
    \includegraphics[width=8cm]{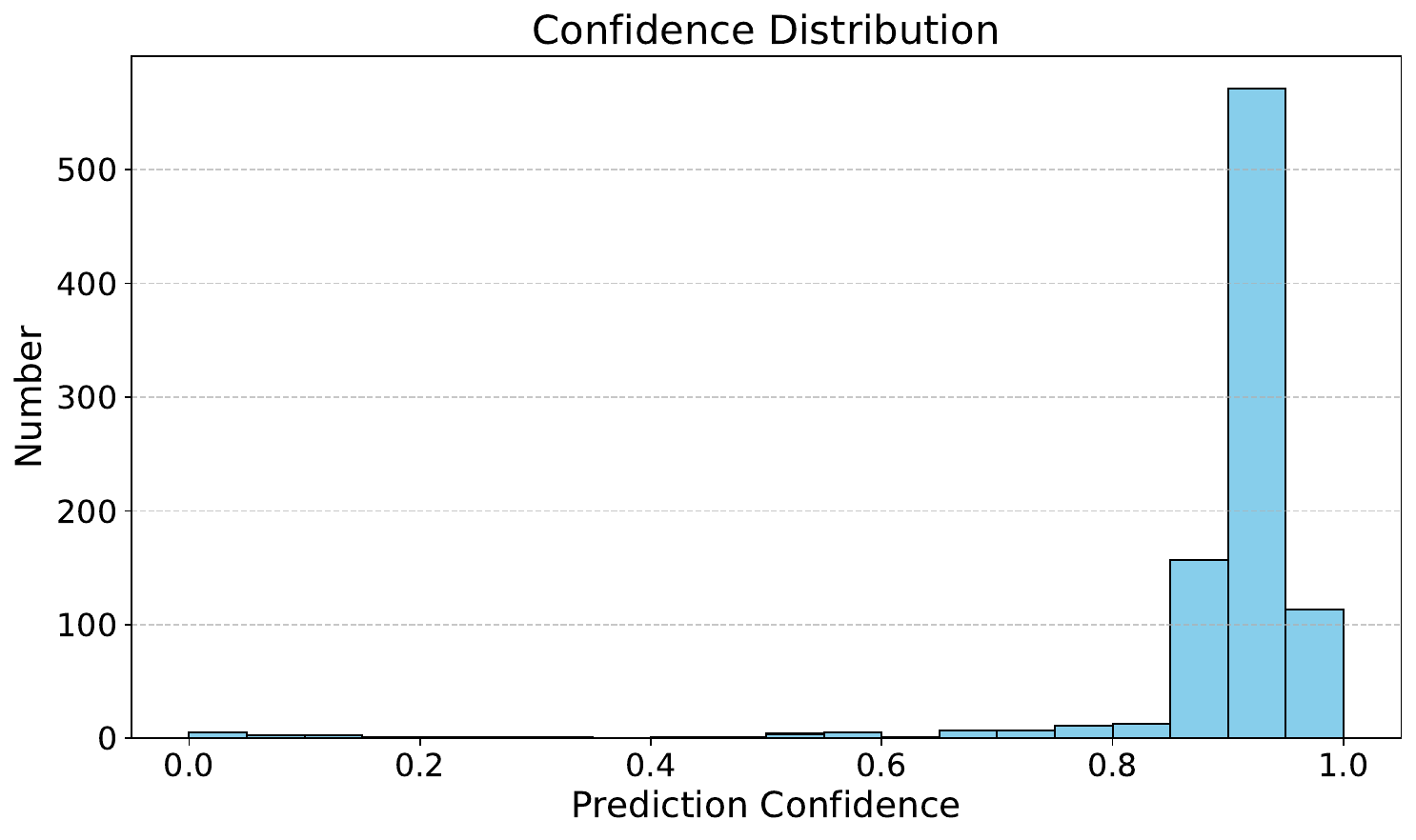}
    \caption{Confidence distribution of LenNet's predicted candidates among the positive samples in the simulated dataset.} 
    \label{fig:confidence_distribution}
\end{figure}

\begin{figure}[h!]
    \begin{center}
        \includegraphics[width=15cm]{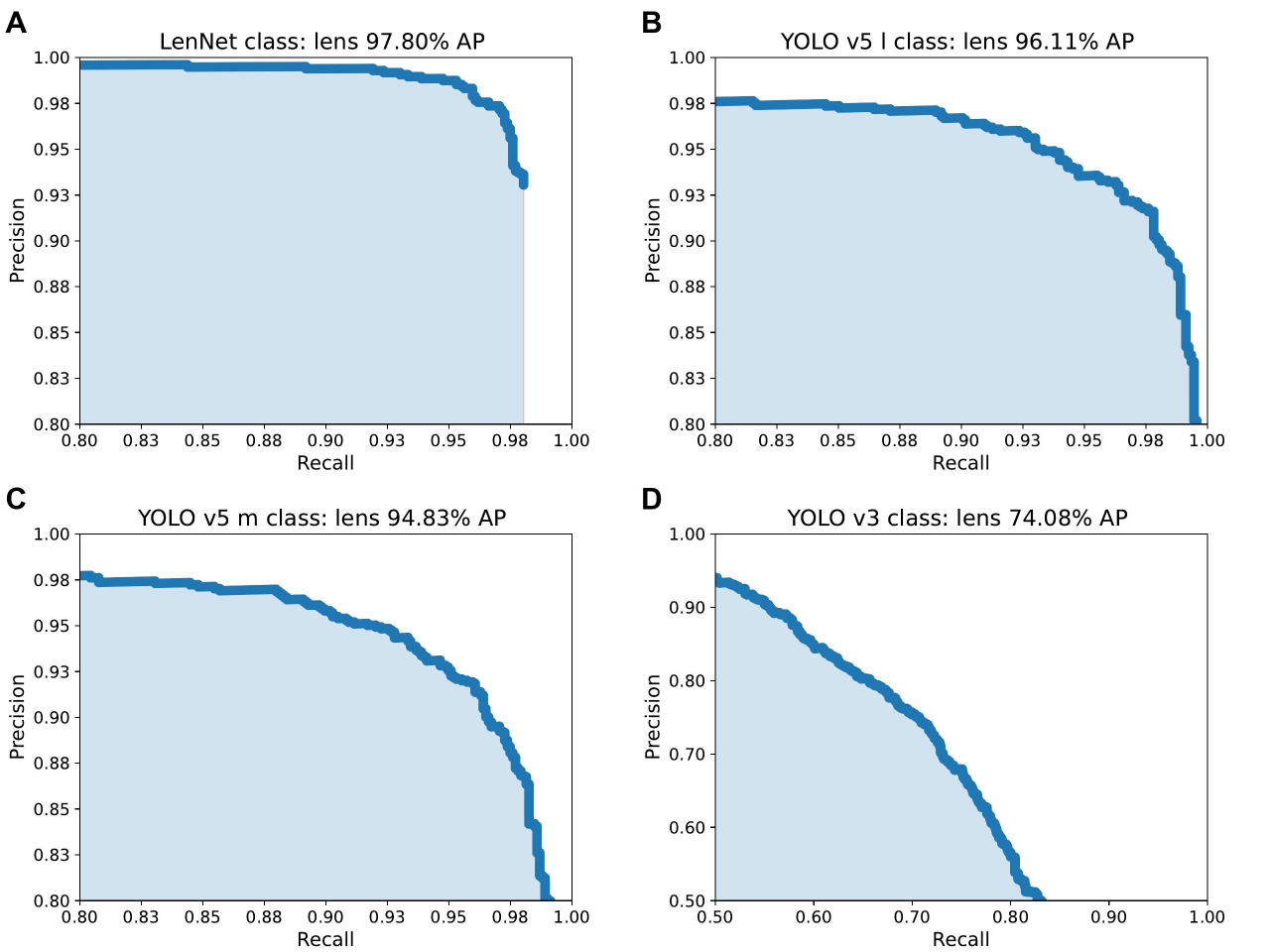}
    \end{center}
    \caption{This figure shows the Precision-Recall curves and Average Precision (AP), which is the area under the precision-recall curve summarizing a model's performance by averaging precision across all recall levels, of four different models on the "lens" category.It is worth noting that the results of YOLO v3 differ significantly from the other three models, so the scale of the coordinates has been adjusted.} 
    \label{fig:Comparison of AP Curves}
\end{figure}

The initial phase of our evaluation focused on quantifying the intrinsic performance of LenNet. On the simulated test set, LenNet attains an F1 score of \(97.11\%\). Evaluating performance across probability thresholds, we set the threshold to 0.3, which yields a recall of \(97.27\%\) and a precision of \(96.95\%\). The high recall indicates that the model successfully identifies nearly all true gravitational lenses, while the high precision demonstrates that its predictions are highly reliable with a low false positive rate. The F1-score, as the harmonic mean of these two, confirms the model's excellent overall balance and accuracy. To supplement these metrics, we conducted an in-depth analysis of LenNet's performance by examining the distribution of prediction confidence scores (Figure~\ref{fig:confidence_distribution}), the Precision-Recall Curves and the Average Precision (Panel A in Figure~\ref{fig:Comparison of AP Curves}). The confidence distribution reveals that the majority of LenNet's predictions have scores exceeding 0.8. That is, most predicted celestial objects have a prediction confidence level above 80\%. Moreover, the AP curve for LenNet is stable and is positioned closest to the top-left corner, achieving an average precision of 96.95\%. This demonstrates the model's excellent precision and recall.

To benchmark LenNet's performance against existing technologies, we conducted a comparative analysis against several models from the YOLO family, which are widely regarded as state-of-the-art in general-purpose object detection. We selected three representative baselines: the classic YOLOv3 and two modern variants, YOLOv5m (medium) and YOLOv5l (large). All models, including LenNet and every network under test, were trained and evaluated under identical experimental conditions on our dataset to ensure a fair and unbiased comparison.The summarized results are presented in Table \ref{tab:sim performance}. The results unequivocally demonstrate that LenNet achieves superior performance across all evaluated metrics. Notably, LenNet's F1-score of 97.11\% surpasses that of the strongest baseline, YOLOv5l (94.52\%), by a significant margin of 2.59 percentage points. This performance gap underscores the primary advantage of a specialized architecture. While the YOLO models are highly optimized for general-purpose object detection, LenNet's design, particularly its MFEBlock modules, is explicitly tailored to discern the subtle and complex morphological features characteristic of gravitational lenses. This domain-specific adaptation allows it to achieve a level of precision and recall that even state-of-the-art generalist models cannot match, thereby validating our architectural design choices and establishing LenNet as a highly effective tool for this specialized scientific task.


\begin{table}[h]
    \centering
    \renewcommand{\arraystretch}{1.0} 
    \begin{tabular}{lccc}
        \toprule
           & F1 & Recall & Precision \\
        \midrule
        LenNet & {97.11\%} & {97.27\%} & {96.95\%} \\
        YOLO v5l & 94.52\% & 97.16\% & 92.03\% \\
        YOLO v5m & 93.88\% & 95.52\% & 92.29\% \\
        YOLO v3 & 68.69\% & 59.23\% & 81.74\% \\
        \bottomrule
    \end{tabular}
    \caption{Comparison of detection performance among different models on the simulated dataset, evaluated using F1-score, recall, and precision. LenNet achieves superior results across all metrics, significantly outperforming other baseline models.}
    \label{tab:sim performance}
\end{table}

\subsection{Transfer learning to the real dataset}\label{sec:4 trans}
Although deep learning models demonstrate excellent performance on simulated data, their scientific utility is ultimately determined by their ability to generalize to real-world observations. A critical challenge in this transition is the ``sim-to-real'' gap, where subtle differences between simulated and real data can degrade performance.
For instance, in tasks like ionized nebula classification, strong gravitational lens substructure detection, and galaxy merger identification, differences between real and simulations in spectral features, image noise, and data distributions have been shown to reduce model accuracy \citep{belfiore2025machine,alexander2023domain,ciprijanovic2021deepmerge}, underscoring the need for techniques like transfer learning. 

\begin{figure}[h!]
    \begin{center}
        \includegraphics[width=15cm]{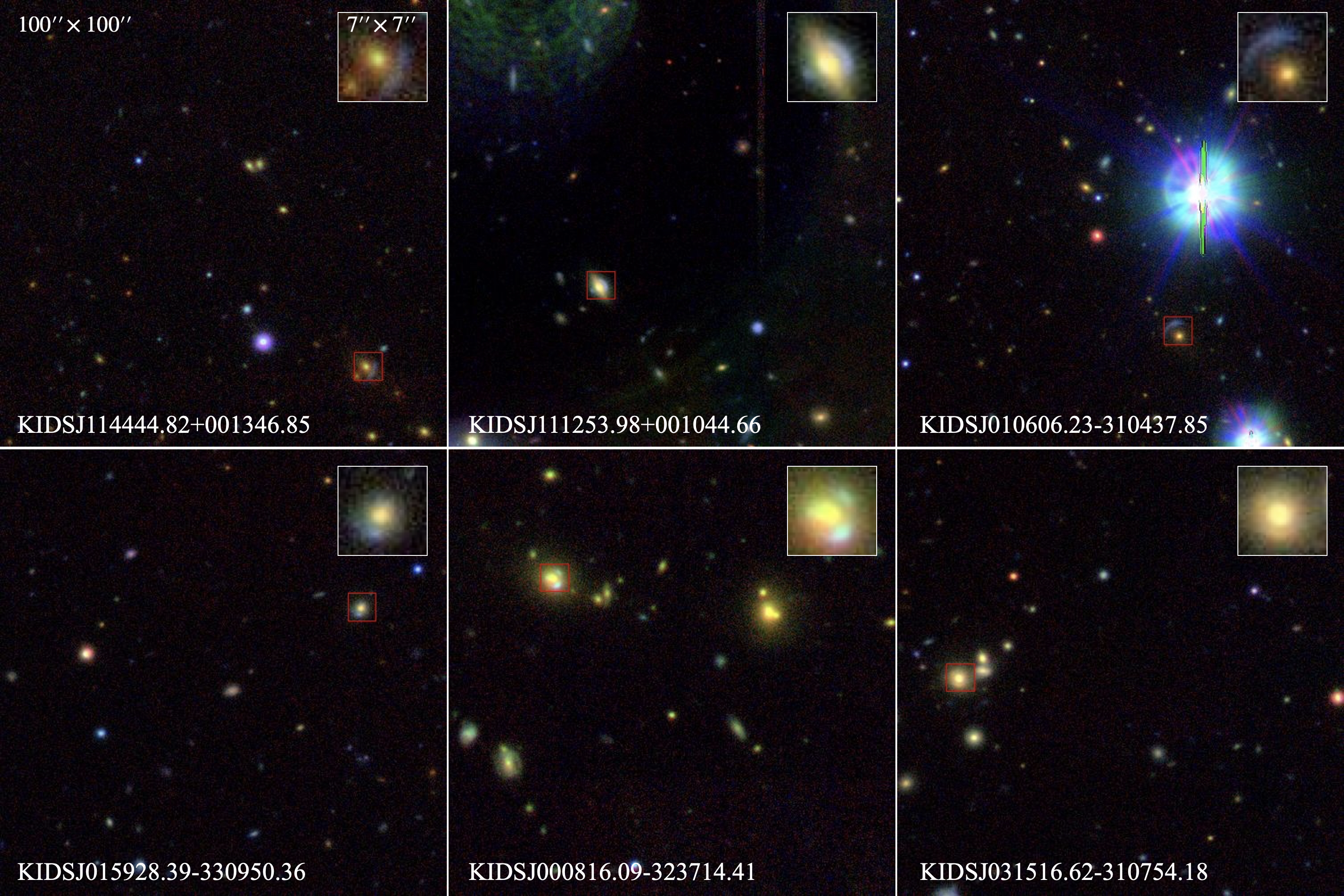}
    \end{center}
    \caption{Examples of gravitational lens candidates from KiDS DR4 used for transfer learning. Each $501 \times 501$ pixel image contains a galaxy-galaxy strong lensing (GGSL) event, with insets highlighting the lensed features.} 
    \label{fig:KIDS true_lenses}
\end{figure}

In this work, the real observations used for transfer learning consist of high-quality lens candidates identified in KiDS DR4 \citep{Petrillo2019, 2020ApJ...899...30L, grespan2024teglie, 2024MNRAS.533.1426N}. Since the criteria for a 'high-quality' candidate vary across these publications, we adopt the classification specified in each respective source.
While most of these candidates lack spectroscopic confirmation, their distinct morphological features in the images suggest a high likelihood of being genuine lenses. For the purpose of testing our networks, we treated these candidates as confirmed lenses. We collected a total of 401 lens candidates, with randomized positions in the $501\times501$ pixel large images as illustrated in Figure~\ref{fig:KIDS true_lenses}. These images were then randomly split into training and testing sets for the transfer learning experiments.To improve the data diversity of the transfer learning dataset, we additionally incorporated non-lens images as supplementary data, with their quantity randomly selected to be the same as the real lenses. For instance, a transfer dataset built on 200 real-world samples will include not only these 200 real samples but also 200 real non-lens images.

To specifically quantify the impact of the “sim-to-real” gap and the necessity of transfer learning, we investigated the performance of LenNet and YOLO models on real images without transfer learning. The test shows that neither the LenNet nor the YOLO model achieves satisfactory performance on real-world images. When tested on 402 real-world images (including 201 with lenses and 201 without lenses), LenNet can only detect 93 targets with 13 false detections when the confidence threshold is set to 0.1. In contrast, under the same confidence threshold of 0.1, even the YOLO v5 l model, which performed best in tests on the simulated dataset, detected only 77 targets while having 73 false detections. These results clearly show the drawbacks of directly applying models trained on simulated data to real-world scenarios, highlighting why transfer learning is indispensable. Furthermore, the scarcity of labelled real-world astronomical data often precludes the reliable training of deep learning models from scratch. We employ transfer learning to address this challenge; this technique leverages knowledge from a model trained on a large, data-rich source domain (our simulations) and adapts it to a target domain with limited data (real observations) \citep{weiss2016survey}. As established in Section~\ref{sec:mock_perform}, the superior feature extraction capabilities of LenNet ensure its pre-trained weights offer a robust foundation for fine-tuning on real images. This approach avoids training a model from scratch on a small dataset---a process prone to overfitting---by instead fine-tuning the existing powerful feature representations to the specific nuances of real survey data.



The transfer learning experiment proceeded as follows: we utilized the LenNet model, pre-trained on a comprehensive simulated dataset, as the initial weights for fine-tuning on real-world data. Benefiting from the pre-trained weights' ability to extract general features, this transfer-learning approach exhibits greater stability. To accelerate model convergence, we moderately increased the learning rate and fine-tuned the model on two small, independent datasets of real-world gravitational lens images. One dataset contains 400 images, consisting of 200 real lens-containing samples and 200 real lens-free samples; the other dataset contains 720 images, consisting of 360 real lens-containing samples and 360 real lens-free samples. Since lens-containing samples are the dominant samples for model training, the 400-image dataset is referred to as the "200-set" and the 720-image dataset as the "360-set" in the following context. To provide a direct performance benchmark, we also applied the identical fine-tuning procedure to YOLOv5m—the strongest baseline from our previous analysis—using the 200-image dataset. The comparative performance metrics are presented in Table~\ref{tab:real performance}. The fine-tuning experiments yield two key insights. Firstly, when both models are fine-tuned on the identical, limited dataset of 200 real images, LenNet significantly outperforms YOLOv5m. Specifically, LenNet achieves an F1-score of 89.47\%, exceeding YOLOv5's score of 81.77\% by about eight percentage points. This outcome strongly suggests that the specialised features learned by LenNet during pre-training are more robust and directly transferable to real-world lens detection. LenNet's architecture, tailored for astrophysical morphologies, provides a more effective starting point, allowing it to adapt more efficiently with limited target data than its general-purpose counterpart. Secondly, the performance of LenNet demonstrates positive scaling with the quantity of training data. When the fine-tuning dataset was expanded from 200-set to 360-set images, the model's F1-score improved from 87.63\% to 89.47\%. This improvement confirms that while LenNet is effective with minimal data, its performance can be systematically enhanced as more labelled examples become available.

In conclusion, these findings validate the effectiveness of our two-stage training strategy (simulation pre-training followed by real-data fine-tuning). The results not only highlight LenNet's strong potential for immediate application in current astronomical surveys but also illuminate a path forward for its continuous improvement. As new gravitational lenses are discovered and validated, they can be incorporated into the fine-tuning set, creating a virtuous cycle where the model's detection capabilities are progressively refined and enhanced over time.

\begin{table}[h]
    \centering
    \begin{tabular}{lcccc}
        \toprule
        \textbf{Model} & \textbf{Number of Images} & \textbf{Recall} & \textbf{Precision} & \textbf{F1 Score} \\
        \midrule
        LenNet & 200 & 84.58\% & 90.91\% & 87.63\% \\
        LenNet & 360 & 82.93\% & 97.14\% & 89.47\% \\
        YOLO v5 & 200 & 82.59\% & 80.98\% & 81.77\% \\
        \bottomrule
    \end{tabular}
    \caption{Performance comparison of the LenNet and YOLOv5 models. Both models were pre-trained on the simulated dataset and subsequently fine-tuned on real lensing images via transfer learning. The second column specifies the number of real lensing images used for the fine-tuning phase.}
    \label{tab:real performance}
\end{table}

\subsection{Observe LenNet from the aspect of feature maps}
\begin{figure}[h!]
    \centering 
    \includegraphics[width=8cm]{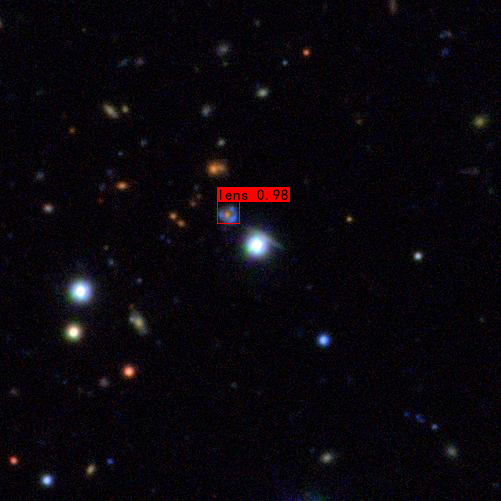}
    \caption{An example of a simulated gravitational lensing image containing an unobscured lens, used to illustrate the interpretability of the \texttt{LenNet} model. The red bounding box indicates the GGSL identified by the model.} 
    \label{fig:feature pre pic}
\end{figure}

Visualizing the model's internal feature maps provides qualitative insight into its decision-making process. This interpretability analysis is critical for verifying that the model learns to identify physically meaningful structures and for fostering confidence in its application as a scientific instrument. For this analysis, we selected a representative test image (Figure~\ref{fig:feature pre pic}) that features an unobscured gravitational lens. Crucially, this image also contains other astronomical objects with morphologies or sizes that could potentially confound the lens finders. We trace the data's progression through the two primary architectural stages of LenNet: the backbone is used for initial feature extraction, and the FPN, which is part of the Feature Fusion Neck, facilitates feature fusion and refinement.

\begin{figure}[h!]
    \centering 
    \includegraphics[width=15cm]{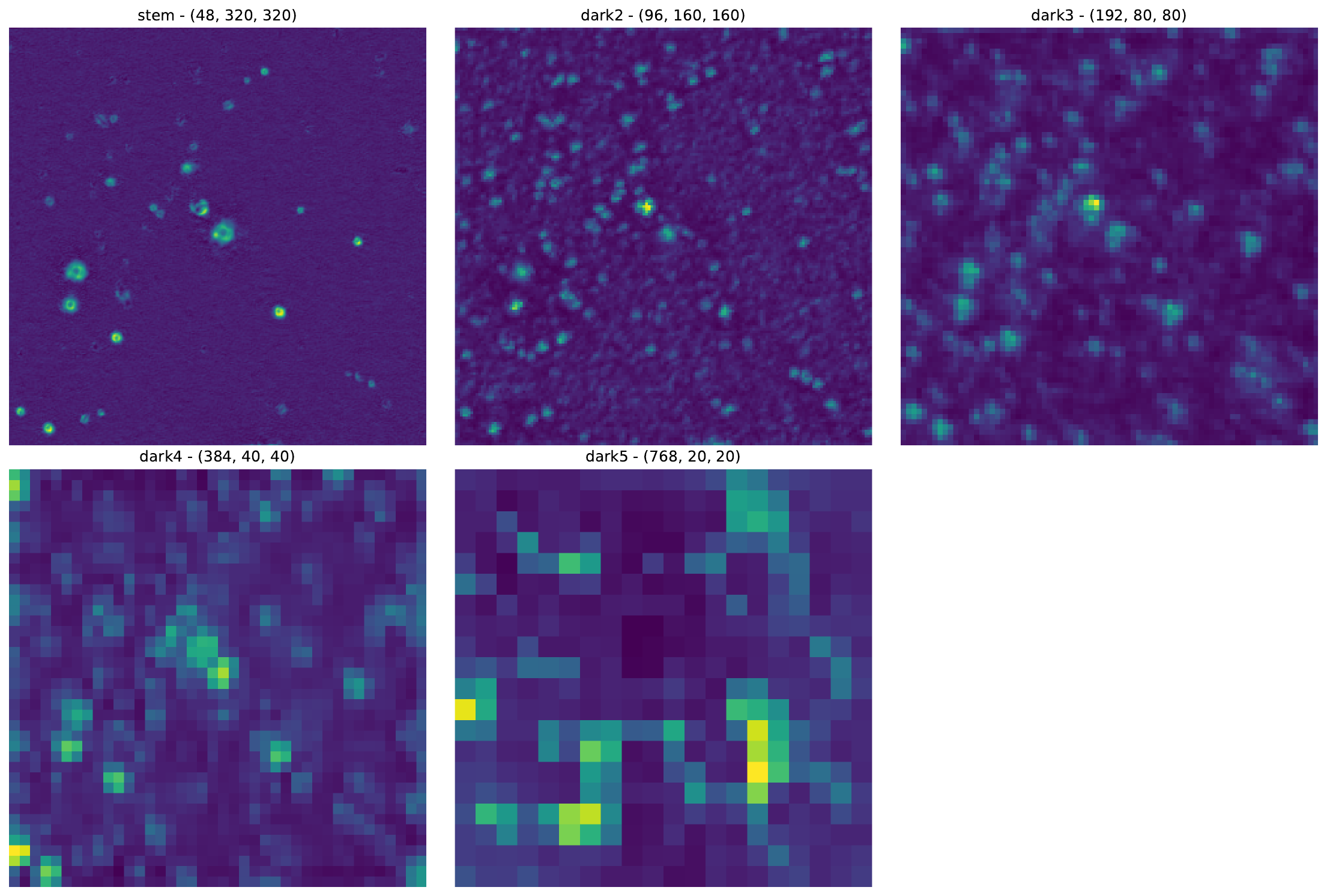}
    \caption{Feature maps from the LenNet backbone for the input survey image presented in Figure~\ref{fig:feature pre pic}.In the image, the blue areas indicate regions where the network does not focus its attention. The 'stem' layer denotes the feature map following initial feature extraction, with 'dark2' to 'dark5' representing feature maps processed by subsequent feature extraction blocks. The tensor shapes of these layers are also illustrated in the model architecture diagram.} 
    \label{fig:backbone_features}
\end{figure}

The backbone serves as the foundational perception module of the network. Its role is to decompose the raw input image into a rich, hierarchical set of feature representations at multiple scales. Figure \ref{fig:backbone_features} displays the feature maps extracted from different layers of the backbone. As observed, the shallower layers capture low-level primitives such as edges, textures, and simple shapes. These activations are distributed across nearly all celestial objects in the frame, indicating that at this stage, the network is performing a broad, class-agnostic analysis of the visual information. As we progress to deeper layers, the features become more abstract and semantically complex, combining the initial primitives into representations of whole objects or significant parts thereof. At this stage, while activations are still present on multiple objects, the network has successfully generated a multi-scale inventory of all potentially salient structures in the image, providing the raw material for the subsequent detection stages.

\begin{figure}[h!]
    \centering 
    \includegraphics[width=15cm]{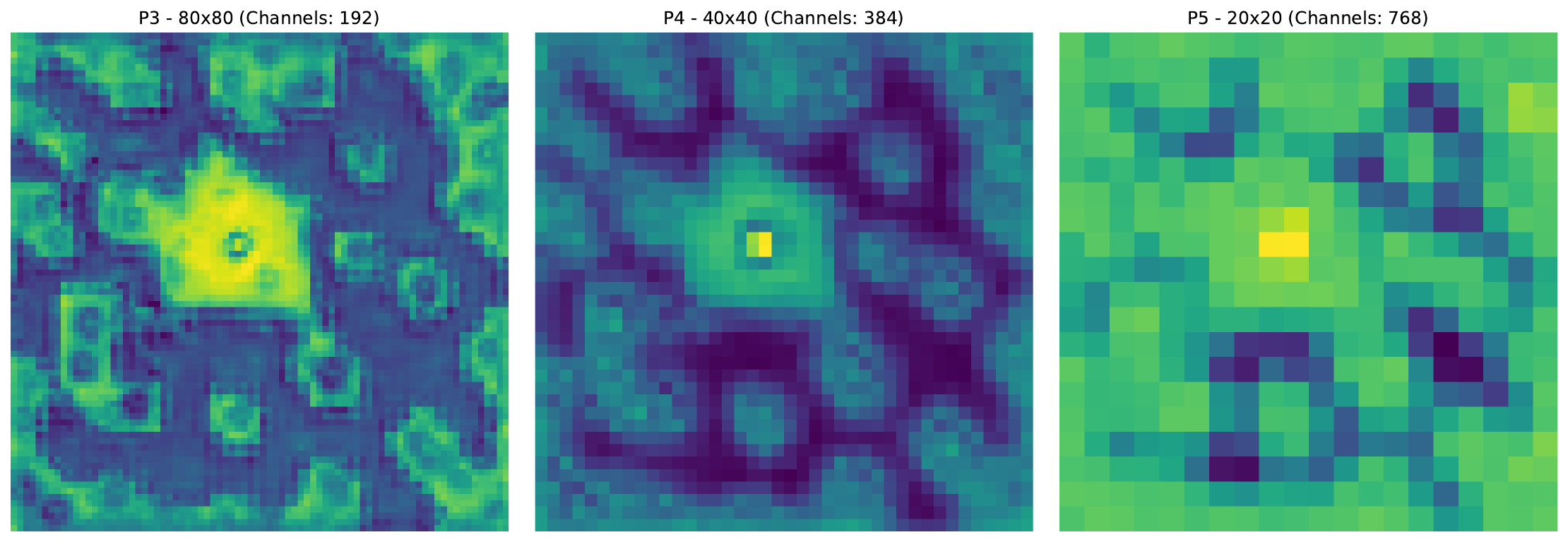}
    \caption{The Feature Pyramid Network (FPN) feature maps of the \texttt{LenNet} model for the input survey image presented in Figure~\ref{fig:feature pre pic}These three images represent the final output of the Feature Fusion Neck, which corresponds to the ultimate feature maps of LenNet. Notably, the yellow regions in the center are highly consistent across the images, suggesting a strong likelihood of the presence of a gravitational lens at these locations. In contrast, the blue areas indicate regions where the model does not focus its attention, implying that it considers these areas unlikely to contain gravitational lenses.} 
    \label{fig:fpn_feature}
\end{figure}

The features extracted by the backbone are then processed by the FPN. The FPN's critical function is to fuse information across different scales, enhancing features relevant to the target class (gravitational lenses) while suppressing background noise and features from irrelevant objects. Figure \ref{fig:fpn_feature} visualizes the output feature maps from the FPN. The transformative effect of the FPN is immediately apparent. A stark contrast emerges between the diffuse activations in the backbone and the highly localized, high-intensity activations in the FPN maps. Across all three scales shown, the features corresponding to the gravitational lens are sharply and decisively enhanced. Simultaneously, the activations on neighboring galaxies and other distractor objects have been significantly attenuated. This visualization powerfully demonstrates that the FPN has learned to function as a highly effective spatial and semantic filter. It successfully integrates the multi-scale information from the backbone to converge on a high-confidence representation of the gravitational lens. This process creates a precise and unambiguous saliency map that provides clear guidance to the final detection head, enabling it to localize the lens with high accuracy. The clarity and focus of these final feature maps directly explain the high precision and recall scores reported in our quantitative analysis.

\section{Conclusion}
\label{sec:conclusion}
This paper presents LenNet, a high-performance model for the automated detection of gravitational lenses. Our results show that LenNet not only excels on simulated datasets but also achieves outstanding performance on real-world data. It significantly outperforms several mainstream detection models, with a recall of 97.27\%, a precision of 96.95\%, and a high F1-score of 97.11\% on simulated test data. Furthermore, feature visualization analysis demonstrates LenNet’s ability to extract key gravitational lens characteristics even in complex astronomical backgrounds, highlighting its robustness and interpretability.

Following its success with simulated data, LenNet maintains strong performance under a transfer learning strategy. The model adapts well to new datasets and requires only a small amount of training data. This approach not only reduces training costs but also preserves high performance, underscoring LenNet’s efficiency and adaptability. Experimental results show that even with just the 200-set, LenNet achieves a recall of 84.58\%, a precision of 90.91\%, and an F1-score of 87.63\%. With the 360-set, the model sees a clear performance boost: recall reaches 82.93\%, precision rises to 97.14\%, and the F1-score improves to 89.47\%. These findings validate the model’s generalization capabilities and demonstrate its practical utility in low-resource scenarios. Compared to training from scratch, transfer learning greatly reduces computational costs, enhancing the model’s scalability.

LenNet clearly exhibits strong potential. With continued training on newly identified lenses, its performance can be further improved. This dynamic adaptability makes LenNet a valuable and evolving tool for gravitational lens detection and astronomical research. Looking ahead, as the volume of sky survey data continues to grow and new samples are accumulated, LenNet can be incrementally optimized through continuous learning. This positions it as a powerful solution for gravitational lens detection in large-scale astronomical imaging. With future deployments, LenNet is expected to support a wide range of practical applications and serve as an important tool in advancing astrophysics and cosmological studies.

\section*{Conflict of Interest Statement}
The authors declare no conflicts of interest regarding the publication of this manuscript.

\section*{Author Contributions}
Pufan Liu built and optimized the \texttt{LenNet} model. Hui Li developed the simulation pipeline to generate the training dataset. Rui Li conceptualized and coordinated the project. Hao Su and Ziqi Li supervised the development of the \texttt{LenNet} model. The original manuscript was written by Pufan Liu, Hui Li, Rui Li, and Xiaoyue Cao, with editing contributions from all authors.

\section*{Acknowledgments}
This work is supported by the National Natural Science Foundation of China (Grant No. 12588202), the China Manned Space Program with grant No.CMS-CSST-2025-A03. Rui Li acknowledges the National Natural Science Foundation of China (No. 12203050). Xiaoyue Cao acknowledges the support of the National Natural Science Foundation of China (No. 12303006). Ran Li is supported by the National Natural Science Foundation of China (Nos. 11773032, 12022306). We acknowledge the use of OpenAI's ChatGPT for assistance with language editing and refinement of this manuscript.

\section*{Data Availability Statement}
The data used in this paper are available from the authors upon reasonable request.

\bibliographystyle{Frontiers-Harvard} %
\bibliography{Manuscript}

\end{document}